\documentstyle[aps,preprint]{revtex}
\begin{document}
\tightenlines
\draft
\title{Fresh inflation: a warm inflationary model 
from a zero temperature initial state}
\author{Mauricio Bellini\footnote{E-mail address: mbellini@mdp.edu.ar}}
\address{Instituto de F\'{\i}sica y Matem\'aticas, \\
Universidad Michoacana de San Nicol\'as de Hidalgo, \\
AP:2-82, (58041) Morelia, Michoac\'an, M\'exico.}

\maketitle
\begin{abstract}
A two -  components mixture fluid
which complies with the gamma law is considered in the framework
of inflation with finite temperature.
The model is developed for a quartic 
scalar potential without symmetry breaking.
The radiation energy density is assumed to be
zero when
inflation starts and remains below the GUT temperature 
during the inflationary stage. 
Furthermore,
provides the necessary number of e - folds and sufficient 
radiation energy density to GUT baryogenesis can
take place 
near the minimum energetic configuration.
\end{abstract}
\vskip 2cm
\noindent
{\rm Pacs:} 98.80.Cq

\section{Introduction and Motivation}

Since its creation, the inflationary Universe scenario\cite{1,2,3,4}
has become an integral part of the standard model of cosmology.
An ideal inflationary scenario should arise naturally from
quantum cosmology\cite{A} without fine tuning.
Inflation is needed because it solves the horizon, flatness, and
monopole problems of the very early universe and also provides
a mechanism for the creation of primordial density fluctuations.
For inflation, naturality has played an important role. This is
understandable since for phenomena that cannot
be directly observed, one attemps a description starting with
the most natural expectations. The importance of naturality
principles is to provide guidance from more familiar analogies
with the hope of faining predictability. For inflation
we can understand naturality as both macroscopic and microscopic.
Macroscopically, we would like a description that rests with
common - day experience. Microscopically, it should be consistent
with the standard model of particle physics.

Quantum fluctuations of matter field\cite{5} and
thermal fluctuations\cite{6} play a prominent role in inflationary
cosmology. They lead to density perturbations that would be responsible
for the origin of structures in the present - day universe\cite{7}.
Structure formation scenarios can receive important restrictions
based on the measured background anisotropy temperature
$\Delta \theta/\theta = 1.1 \  10^{-5}$.

A successful model of inflation must satisfy the following requirements.

a) An unstable primordial matter - field - perturbation 
with a energy density nearly $M^4_p$ must lead to inflation.
b) The growing of the scale factor during inflation must grow at least
by 60 e-foldings, to solve the horizon problem
to give a sufficiently globally flat universe.
c) Inflation must generate density perturbations of temperature with
amplitude constrained by the Cosmic Microwave Background (CMB)
density fluctuations.
d) During all the inflationary era, the temperature must remains
below the GUT critical temperature $\theta_{GUT} \simeq 10^{15}$ GeV,
to avoid monopole and domain walls proliferation.
e) At the end of inflation the universe must be
with sufficient radiation energy density to make posible the
particles creation in the framework of GUT baryogenesis.

The standard slow - roll inflation model separates expansion and reheating
into two distinguished time periods. It is first assumed that
exponential expansion from inflation places the universe in a supercooled
second order phase transition. Subsequently thereafter the universe is
reheated. Two outcomes arise from such a scenario. First,
the required density perturbations in this cold universe are left to be
created by the quantum fluctuations of the inflaton. Second, the
scalar field oscillates near the
minimum of its
effective potential and produces elementary particles. These
particles interact with each other and eventually they come
to a state of thermal equilibrium at some temperature $\theta$. This process
completes when all the energy of the classical scalar field
transfers to the thermal energy of elementary particles.
The temperature of the universe at this stage is called
the reheating temperature\cite{kls}.

The issue of reheating after inflation in these theories, where
the universe is revived from the frozen vacuum state, has remained
relatively unexplored. This is precisely one
area where the full symmetry and particle content of the
underlying theory is likely to be crucial\cite{8a}.
For GUT baryogenesis the large quantum fluctuations may
cause the monopole and domain wall problems due to
nonthermal symmetry restoration\cite{8}.
Hence, if all the particles are
created by means of this mechanism at the end of inflation, the
universe could comes very inhomogeneous.
Furthermore, reheating requires globally coherent radiation waves on
the scale of the inflated universe. A globally coherent heating
process requires a large - scale radiator, which in standard inflation
scenario is the inflaton. The question here is how the random
inflaton field configuration before and during inflation attains
quantum coherence at the end of inflation.

On the other hand
the warm inflation scenario takes into account separately, the matter and
radiation energy fluctuations. In this scenario the fluctuations of the
matter field lead to perturbations of matter and radiation energy
densities\cite{9}, which are responsible for the fluctuations
of temperature. In this scenario the matter field interacts with
particles which are in a thermal bath with
a mean temperature smaller than the GUT critical temperature.
This scenario was introduced by Berera\cite{10}.
Other approaches
for this scenario also were developed\cite{9,11}.
In principle, a permanent or temporary coupling of the
scalar field $\phi$ with other fields might also lead
to dissipative processes producing entropy at different
eras of the inflacionary stage.
The warm inflation
scenario appears to be very promising, but the problem with it
is that, initially, the thermal bath is unjustified introduced
in the framework of chaotic initial conditions needed to give
a natural beining to the universe. In this sense, chaotic inflation\cite{LL}
provides a more natural scenario to describe the initial conditions in the 
universe.
A successful model of inflation for an inflaton that interacts with particles
in a thermal bath initially must be with
$\rho_r(t=t_o) =0$ because the universe is, in principle,
created from a quantum fluctuation of the vacuum.
In this framework, the temperature of the universe
must be increasing during inflation to baryogenesis can take place.

The aim of this work is to propose a model of inflation
that incorporates some characteristics of both,
standard (chaotic) and warm inflationary scenarios to give
a natural inflationary scenario in the framework of chaotic initial 
conditions.

\section{General considerations}

I consider an homogeneous and isotropic universe, described by a flat
Friedmann - Robertson - Walker (FRW) metric
\begin{equation}\label{1}
ds^2 = - dt^2 + a^2(t) dr^2,
\end{equation}
where $a(t)$ is the scale factor of the universe. The early
universe can be represented by the Lagrangian density
\begin{equation}\label{2}
{\cal L} = \frac{1}{2} \partial^{\mu} \phi \partial_{\mu} \phi -
V(\phi) + {\cal L}_{\rm int},
\end{equation}
where ${\cal L}_{\rm int}\sim -g^2 \phi^2 \Psi^2$ 
describes the interaction between the
scalar field $\phi$ and the other $\Psi$-scalar fields 
of a thermal bath.
The minimally coupled scalar field and the
fields of a thermal bath interchange energy during the
rapid expansion of the universe.
The various outcomes are a result of specially chosen
Lagrangians. In the most of cases the Lagrangian is unmotived from
particle phenomenology. Clear exceptions are the Coleman - Weinberg
potential with coupling constant which arise from GUT theories
and SUSY potentials\cite{!}.

The Einstein's equations for this system are
\begin{eqnarray}
3 H^2 & = & 8 \pi G \left[ \frac{\dot\phi^2}{2} +
V(\phi) + \rho_r \right], \label{3} \\
3 H^2  + 2 \dot H & = & - 8 \pi G \left[ \frac{\dot\phi^2}{2}
- V(\phi) + \frac{\rho_r}{3}\right], \label{4}
\end{eqnarray}
where $H = {\dot a \over a}$ is the Hubble parameter, $G = M^{-2}_p$ is the
gravitational constant, $M_p = 1.2 \times 10^{19}$ GeV is the Planckian
mass and $\rho_r$ is the radiation energy density. 
The equation of motion for 
$\phi$ in an interacting system is
\begin{equation}\label{6}
\dot\phi \left[ \ddot\phi + 3 H \dot\phi + V'(\phi)\right]
= -\delta,
\end{equation}
where $\delta = \dot\rho_r + 3 \gamma H \rho_r$ describes the interaction
between both subsystems; the inflaton field and the bath.
Furthermore, $\rho_r$ is the radiation energy density.
The eq. (\ref{6}) can be written as two equations
\begin{eqnarray}
&&\ddot\phi
+ 3 H \dot\phi + V'(\phi) + \frac{\delta}{\dot\phi} =0, \label{7}\\
&&\dot\rho_r + 3\gamma H \rho_r - \delta = 0, \label{8}
\end{eqnarray}
where $\gamma$ takes the value $4/3$ in warm inflationary models.
Some phenomenological interaction terms as
$\delta \propto \dot\phi^2$\cite{10},
$\delta \propto \phi^2 \dot\phi^2$\cite{11}
or $\delta \propto \dot\phi^d \phi^{5-2d}$\cite{12}, has been
proposed in the literature.
Very much work
has been made in this framework for a de Sitter expansion of the
universe\cite{13}. However, more interesting and complicated is to work
in a quasi - de Sitter expanding universe\cite{9},
where the fluctuations of the matter field are considered.
One of the more interesting consequences that
arise from these fluctuations
are the metric fluctuations\cite{14} on a FRW background metric.

Slow-roll conditions must be imposed to assure nearly de Sitter
solutions for an amount of time, which must be long enough to solve
the problems of the hot big bang.
If $p_t = {\dot\phi^2\over 2} + {\rho_r \over 3} -V(\phi)$ 
is the total pressure and
$\rho_t = \rho_r + {\dot\phi^2 \over 2} + V(\phi)$ 
is the total energy density, hence
the parameter $F = {p_t+ \rho_t \over \rho_t}$ which describes slow
roll conditions is\cite{rc}
\begin{equation} \label{F}
F =- \frac{2 \dot H}{3 H^2} = \frac{\dot\phi^2 + \frac{4}{3} \rho_r}{
\rho_r + \frac{\dot\phi^2}{2} + V}.
\end{equation}
The requirement to assure slow-roll conditions is that
$F \ll 1$ and, from eq. (\ref{F}) one obtains the following
equations
\begin{eqnarray}
&& \dot\phi^2 \left(1-\frac{F}{2} \right) + \rho_r \left(
\frac{4}{3} - F\right) - F V(\phi) =0, \label{a1}\\
&& H= \frac{2}{3 \int F dt}. \label{a2}
\end{eqnarray}
Furthermore, due to $\dot H = H' \dot\phi$, from the first
equality in (\ref{F}) we obtain the time dependence for
$\phi$ 
\begin{equation}\label{dot}
\dot\phi = - \frac{3 H^2}{2 H'} F,
\end{equation}
such that from eqs. (\ref{a1}) and (\ref{dot})
the radiation energy density can be written
as a function of $V(\phi)$ and the Hubble parameter
\begin{equation}\label{b1}
\rho_r = \left(\frac{3F}{4-3F}\right) V - \frac{27}{8}
\left(\frac{H^2}{H'}\right)^2 \frac{F^2 (2-F)}{(4-3F)}.
\end{equation}
Furthermore, replacing (\ref{dot}) and (\ref{b1}) in eq. (\ref{3}), one
obtains the scalar field potential as a function of $H$, for a given
parameter $F$ 
\begin{equation}\label{V}
V(\phi) = \frac{3}{8\pi G}
\left[ \left(\frac{4-3F}{4}\right) H^2 +
\frac{3\pi G}{2} F^2 \left(\frac{H^2}{H'}\right)^2\right].
\end{equation}
If $F \ll 1 $ is a constant, the evolution for the Hubble parameter 
$H= {\dot a \over a}$ being given from eq. (\ref{a2})
\begin{equation}\label{14}
H(t) = \frac{2}{3 F} t^{-1}, 
\end{equation}
such that 
\begin{equation}
a(t) \sim t^{\frac{2}{3F}}.
\end{equation}
Finally, the  number of e-folds $N=\int^{t_e}_{t_s} H dt$ ($t_s$ and
$t_e$ are the time when inflation start and ends), is given by
\begin{equation}
N = \left.\frac{2}{3F} {\rm ln}(t)\right|^{t_e}_{t_s}.
\end{equation}
With Planckian unities ($G^{-1/2}\equiv M_p =1$) 
inflation starts when 
$t_s=G^{1/2}=1$.
Hence, for $t_e \simeq 10^{10} \  G^{1/2}$, one obtains
$N > 60$ for $F<0.55$.

\section{The model}

The idea of fresh inflation
is that the universe is reheating during inflation and
the temperature always remains below $\theta_{GUT}$.
This point is very important to avoid the magnetic monopole
restoration. 
If when inflation starts the inflaton field is not
thermalized, the radiation component of energy density
must be zero at this moment, so that $\rho_r(t_s)=0$. 
For a model with finite temperature, the effective
potential must be dependent of the temperature $\theta=K T$ (here
$K$ is the Boltzmann constant). 

\subsection{The effective potential}

I will consider a nongauge theory, invariant under
a global group {\it O(n)}, involving a single $n$-vector multiplet of scalar
fields $\phi_i$. At zero temperature it is
\begin{equation}\label{15}
V(\phi_i) = \frac{{\cal M}^2_o}{2} \phi_i \phi_i +
\frac{\lambda^2}{4} \left(\phi_i \phi_i\right)^2, 
\end{equation}
where ${\cal M}^2_o > 0$ and $\lambda^2 >0$. Note that I am considering
the model without symmetry breaking.
With the notation $\left(\phi_i \phi_i\right)^{1/2}\equiv \phi$, the
effective finite temperature potential $V_{eff}(\phi,\theta)=
V(\phi) + \rho_r (\phi,\theta)$ can be written as
\begin{equation}\label{18}
V_{eff}(\phi,\theta) = \frac{{\cal M}^2(\theta)}{2} \phi^2 +
\frac{\lambda^2}{4} \phi^4 ,
\end{equation}
where 
\begin{equation}\label{19}
{\cal M}^2(\theta)  = 
{\cal M}^2(0) + \frac{(n+2)}{12} \lambda^2 \theta^2.
\end{equation}
Here, $V_{eff}(\phi,\theta)$ is invariant under $\phi \rightarrow - \phi$
reflexions and $n$ is the number of created particles due to the interaction
of $\phi$ with the fields of the bath.
Furthermore, ${\cal M}^2(0)$ 
is given by ${\cal M}^2_o$ plus
conunterterms\cite{Weinberg}.

From eq. (\ref{V}) one obtains the Hubble parameter as a function of
$\phi$
\begin{equation}\label{20}
H(\phi) = 4 \sqrt{\frac{\pi G}{3(4-3 F)}} \  {\cal M}(0) \  \phi,
\end{equation}
and the condition of consistence implies that
\begin{equation}
\lambda^2 = \frac{12 \pi G}{(4-3 F)} \  {\cal M}^2(0).
\end{equation}
Since $V_{eff}(\phi,\theta) = V(\phi) + \rho_r$ and $\rho_r =
{\pi^2 \over 30} g_{eff} \theta^4$, from eqs. (\ref{18}) and (\ref{19})
one obtains the equality 
\begin{equation}
\frac{(n+2)}{12} \lambda^2 \theta^2 \phi^2=\frac{\pi^2}{30} g_{eff} \theta^4,
\end{equation}
such that
the number of created particles
during inflation will be
\begin{equation}
(n+2) = \frac{2\pi^2}{5 \lambda^2} g_{eff} \frac{\theta^2}{\phi^2},
\end{equation}
where $g_{eff}$ is the effective degrees of freedom for the particles.
I consider a Yukawa interaction like\cite{10}
\begin{equation}\label{23}
\delta(\dot\phi,\phi,\theta) = \Gamma(\theta) \  \dot\phi^2,
\end{equation}
where\cite{BR1,yl}
\begin{equation}\label{gamma}
\Gamma(\theta)  = \frac{g^4_{eff}}{192 \pi} \theta, 
\end{equation}
is the decay width of the particles. Furthermore, 
the thermal bath is at temperature
$\theta \sim \rho^{1/4}_r$ assuming, that $\Psi$ has no self-interaction.
The most
important difference between standard and warm inflation is that
the slope of inflationary does not need to be small, as dissipation
will allow slow - roll. 
From eqs. (\ref{14}) and (\ref{20}) we can find the temporal evolution
for $\phi$
\begin{equation}\label{phi}
\phi(t) = \lambda^{-1} \  t^{-1},
\end{equation}
such that, if inflation starts at $t_s = G^{1/2}$ and
ends at $t_e = 10^{10} \  G^{1/2}$, hence the respective values for 
the field will be $\phi_s \simeq 10^7 \  G^{-1/2}$ and
$\phi_e \simeq 10^{-3} \  G^{-1/2}$, for $\lambda^2 =10^{-14}$. 
Replacing (\ref{gamma}) in (\ref{23}), and using the eqs.
(\ref{phi}), (\ref{14}) and (\ref{8}), we find the
temperature as a function of time
\begin{equation}
\theta(t) = \frac{192 \pi}{g^4_{eff} \lambda^2 }
\left\{ {\cal M}^2(0)\lambda^2 t+ t^{-1}
\left[\lambda^2\frac{(9F^2-18 F+8)}{(4-3F)^2}+{\cal M}^2(0) \pi G 
\frac{(192 F^2
-72 F^3 - 96)}{(4-3F)^2}\right] \right\},
\end{equation}
which increases linearly with time for $t\gg 1$.
If we take $\theta_e =0.1 \  \theta_c \simeq 10^{-4} \  G^{-1/2}$, 
the number of created
particles at the end of inflation is of the order of
(I am taking $g_{eff}=10^2$)
\begin{equation}
n_e \simeq 4 \times 10^8.
\end{equation}

\subsection{Dynamics for the inflaton field and density fluctuations}

In this section I will study the dynamics for the inflaton field
to make an estimation for the energy density fluctuations in a
globally flat Friedmann - Robertson - Walker (FRW) metric
\begin{equation}
ds^2 = -dt^2 + a^2(t) dx^2.
\end{equation}
The dynamics for the spatially homogeneous inflaton field is given
by
\begin{equation}\label{sf}
\ddot\phi + \left(3 H + \Gamma \right) \dot\phi + V'(\phi) =0,
\end{equation}
where 
$V'(\phi) \equiv {dV\over d\phi}$. 
The term $\Gamma \dot\phi$ is added in the scalar field equation
of motion (\ref{sf}) to describe the continuous energy transfered from
$\phi$ to the radiation field. This persistent thermal contact during
warm inflation is so finely adjusted that the scalar field evolves always
in a damped regime.

Furthermore, the fluctuations of the field $\delta\phi(\vec x,t) 
\equiv \psi(\vec x,t)$ are
described by the equation of motion
\begin{equation}\label{e}
\ddot\psi - \frac{1}{a^2} \nabla^2\psi + \left(3H + \Gamma\right) \dot\psi
+ V''(\phi) \psi =0.
\end{equation}
Here, the additional second term appears because the fluctuations $\psi$
are spatially inhomogeneous. 
The eq. (\ref{e}) can be mapped in a Klein - Gordon one, making
$\chi = \psi \  e^{3/2\int \left(H+{\Gamma\over 3}\right) dt}$
\begin{equation}\label{f}
\ddot\chi - \frac{1}{a^2} \nabla^2\chi - \frac{k^2_o}{a^2}\chi = 0,
\end{equation}
where $k_o$ is the wavenumber that separates the ultraviolet and
infrared sectors, such that
\begin{equation}
k^2_o = a^2 \left[ \frac{9}{4}\left(H+\frac{\Gamma}{3}\right)^2
+ 3 \left(\dot H + \frac{\dot\Gamma}{3}
\right) - V''[\phi(t)]\right].
\end{equation}
The redefined coarse-grained matter field fluctuations
$\chi_{cg} = \psi_{cg} e^{3/2\int(H+\Gamma/3)dt}$, can be written
as a Fourier expansion of the form
\begin{equation}
\chi_{cg} = \frac{1}{(2\pi)^{3/2}} {\Large\int} d^3k \  \theta(\epsilon k_o-
k) \  \left[a_k e^{i \vec k.\vec x} \xi_k(t) + a^{\dagger}_k
e^{-i\vec k.\vec x} \xi^*_k\right],
\end{equation}
where the asterisk denotes the complex conjugate and $\epsilon \ll 1$ is a
dimensionless parameter.  However, classicality conditions require that
the modes for $k < \epsilon k_o$ be real\cite{CQG99}.
The requirement to inflation holds is that $k^2_o >0$.
The infrared sector describes the spectrum with wavelengths much
bigger than the size of the horizon. This sector is unstable and it is
very important because  the spatial inhomogeneities in this sector
should be the responsible of the structure formation during the 
matter dominated era of the universe.

To calculate $k^2_o$ we need to know
the temporal evolution of
$H(t)$, $V''[\phi(t)]$ and $\Gamma[\theta(t)]$, which are
\begin{eqnarray}
H[\phi(t)] & = & \frac{2}{3 F} t^{-1}, \\
V''[\phi(t)] 
& = &  {\cal M}^2(0) + 3 t^{-2},\\
\left.\Gamma[\theta(t)]\right|_{t \gg 1} & \simeq & {\cal M}^2(0) \  t.
\end{eqnarray}
Hence,
the parameter of mass $\mu^2(t) =
{k^2_o \over a^2}$ at the end of inflation
will be 
\begin{equation}
\left.\mu^2(t)\right|_{t\gg 1} \simeq  
\frac{{\cal M}^4(0)}{4} t^2 + \left(\frac{1}{F^2}-3\right) t^{-2} +
{\cal M}^2(0) \left(\frac{1}{F}-\frac{1}{2}\right),
\end{equation}
which must be positive during inflation.
The equation of motion for the temporal zero-mode of $\chi$ is
\begin{equation}\label{psi}
\ddot\xi_0  - \left.\mu^2(t)\right|_{t\gg 1} \xi_0(t) \simeq 0,
\end{equation}
which has the general solution
\begin{eqnarray}
\xi_0(t) &=& 
C_1 \sqrt{\frac{t_0}{t}} \  {\rm M}\left[\frac{F-2}{4F},
\frac{\sqrt{4-F(4+11F)}}{4F},\frac{{\cal M}^2(0)}{2} t^2\right] \nonumber \\
&+&
C_2 \sqrt{\frac{t_0}{t}} \  {\rm W}\left[\frac{F-2}{4F},\frac{
\sqrt{4-F(4+11F)}}{4F},\frac{{\cal M}^2(0)}{2} t^2\right], 
\end{eqnarray}
where ($C_1,C_2$) are arbitrary constants and (${\rm M},
{\rm W}$) are the Whittaker functions. 
The asymptotic super Hubble zero-modes matter field fluctuations
are (taking $C_1=0$)
\begin{equation}
\left<\psi^2_{cg}\right> \simeq \frac{C^2_2 \epsilon^3}{6\pi^2} \left(\frac{
t_0}{t}\right) \mu^3(t) \  e^{-\frac{{\cal M}^2(0)}{2} t^2} \  
\left[{\rm W}\left[\frac{F-2}{4F},\frac{\sqrt{4-F(4+11F)}}{4F},
\frac{{\cal M}^2(0)}{2} t^2\right]\right]^2.
\end{equation}
Note that $\left<\psi^2_{cg}\right>$
decreases with time for $t \gg 1$.
The expression for the energy density fluctuations ${\delta\rho_t\over \rho_t} 
\simeq {V'_{eff}\over \dot\phi^2 + {4\over 3} \rho_r}$ at the end
of inflation can be written as a function of $\phi$ and $\theta$
because $\phi(t) = \lambda^{-1} t^{-1}$
\begin{eqnarray}
\left.\frac{\delta\rho_t}{\rho_t}\right|_{end} &\simeq & 
\frac{C_2 \epsilon^{3/2}}{\sqrt{6} \pi} \sqrt{\frac{\phi}{\phi_s}}
\mu^{3/2}(\phi) e^{-\frac{{\cal M}^2(0)}{4\lambda^2}\phi^{-2}}
\left(\frac{{\cal M}^2(0) \phi + \lambda^2 \phi^3}{\lambda^2\phi^4+
\frac{4\pi^2}{90}g_{eff} \theta^4}\right) \nonumber \\
& \times & \left.{\rm W}
\left[\frac{F-2}{4F},\frac{\sqrt{4-F(4+11F)}}{4F},
\frac{{\cal M}^2(0)}{2} \phi^{-2}\right]\right|_{\phi_e,\theta_e},
\end{eqnarray}
where, as was previously calculated, the values at the end of inflation
for the matter field and the temperature are $\phi_e\simeq 0.001 \  G^{-1/2}$
and $\theta_e \simeq 10^{-4} \  G^{-1/2}$.
Using $\epsilon =10^{-3}$ and $F=0.3$, one can calculate the
value for the constant $C_2$ to obtain the experimental COBE data
amplitude for the fluctuations at the end of inflation:
${\delta\rho_t \over \rho_t} \simeq 1 \times 10^{-5}$
\begin{equation}
C_2 \simeq 2.21 \times  10^{12},
\end{equation}
where the calculations were made using ${\cal M}^2(0) 
= 10^{-14} \  G^{-1}$.

\section{Final Comments}

A possible negative aspect of other warm inflationary models
is closely related to a possible thermodynamic fine-tunning, because
an isothermal evolution of the radiative component is assumed from the
very beginning. In the model here developed 
inflation begins from zero radiation energy density.
This is the
more significative difference with other warm inflationary models.
Note than when inflation starts the model gives $\Gamma \ll H$, but nearly
the equilibrium configuration (i.e., for $t \gg 10^7 \  G^{1/2}$)
the situation is inverse [$H(\phi_e) \ll \Gamma(\phi_e,\theta_e)$],
and the thermal equilibrium
is guaranted. Such an effective friction paramter gives
a radiation dominated phase at the end of fresh inflation [$\rho_r(\theta_e) 
\simeq (10^{-3} \  G^{-1/2})^4$], which is needed for a natural
warm inflation - hot big bang transition.

The main difference between standard inflation and the model here
studied is that here the
universe is heating during the inflationary regime. Furthermore, the
dynamics of this heating being perfectly described.
Reheating is not a minor phase at the end of standard inflation.
Standard inflation may cause the monopole and domain
wall nonthermal symmetry
restoration\cite{8}, which could give
a very inhomogeneous universe that disagree with observation.
This last possibility is actually
rather difficult in simple models of reheating with only
two fields\cite{bo}. However, this situation can be solved
in the case of multiple fields, relevant for GUT models\cite{ba}.
This is the case here studied, where the number of fields is nearly
$n \simeq 4 \times 10^{8}$ for the set of valued parameters of the model.
In the model here studied, the particle creation can be justified
by means of the interaction between the
inflaton with other particles of the
thermal bath\cite{10,13,14}.
However, the main difference between this model and standard inflation resides 
in that here there is not oscillation of the inflaton field around
the minimum of the potential, due to dissipation is too large at the
end of inflation. Here, particles creation becomes during inflation, begining
it at zero temperature. Recently, Chung {\em et. al.}\cite{Chung} 
showed that resonant
production of particles during inflation from a zero temperature initial
state can take place. More recently, A. Berera and R. O. Ramos\cite{BR} 
demostrated
that the zero temperature initial state constitutes a baseline effect that
should be revalent in any general statistical state, which gives
a strong evidence that dissipation is the norm not exception 
for an interacting scalar field system. However, must be noted that
this formalism is realized in an non-expanding Minkowski spacetime.
In this work I attempted to develope a model 
in which the universe, starting from chaotic initial conditions, expands
in an increasing damped regime product of the interaction of the
inflaton field with other scalar fields of a zero temperature
initial state. The 
temperature increases with the expansion of the universe
because the inflaton transfers radiation energy density to the bath
with rate bigger than the expansion of the universe. But the crucial
point here is that this model attempt to build a bridge between standard
and warm inflationary models, begining from chaotic initial conditions
which provides naturality. This is the main difference between the
scenario here worked with other.
Finally, the model here studied
may provides the necessary number of e-folds
to explain the flatness/horizon
problem for ${p_t+\rho_t \over \rho_t} < 0.55$. 
This inequality assures slow-roll conditions during
the inflationary expansion of the universe.
The
model provides
enough postinflationary radiation temperature
at the end of inflation to baryogenesis can take place after inflation.
Furhtermore,
the amplitude for energy density fluctuations ($\delta \rho_t/\rho_t$)
decreases with time, being of the order of ${\delta\rho_t \over \rho_t}
\simeq 1 \times 10^{-5}$ when $t$ assumes the value
$t_e =10^{10} \  G^{1/2}$.\\
\vskip .2cm
\centerline{ACKNOWLEDGMENTS}
\vskip .2cm
\noindent
MB acknowledges the support of CONACYT (M\'exico).\\

\end{document}